\documentclass[fleqn,usenatbib]{mnras}

\usepackage{newtxtext,newtxmath}

\usepackage[T1]{fontenc}
\usepackage{ae,aecompl}

\usepackage{graphicx}	
\usepackage{amsmath}	
\usepackage{amssymb}	
\usepackage{xcolor}
\usepackage{cancel}
\usepackage[export]{adjustbox}
\usepackage{gensymb}

\title[Water maser superburst in G25.65+1.05]{VLBI observations of the G25.65+1.05 water maser superburst}

\author[R. A. Burns et al.]{
R. A. Burns,$^{1,2,3}$\thanks{E-mail: ross.burns@nao.ac.jp}
G. Orosz,$^{4,5}$
O. Bayandina,$^{1,6}$
G. Surcis,$^{7}$
M. Olech,$^{8}$
G. MacLeod,$^{9,10}$\newauthor 
A. Volvach,$^{6,11}$
G. Rudnitskii,$^{12}$
T. Hirota,$^{2}$
K. Immer,$^{1}$
J. Blanchard,$^{1,13}$
B. Marcote,$^{1}$\newauthor 
H. J. van Langevelde,$^{1,14}$
J. O. Chibueze,$^{15,16}$ 
K. Sugiyama,$^{2,17}$
Kee-Tae Kim,$^{3,18}$\newauthor
I. Val`tts,$^{6}$
N. Shakhvorostova,$^{6,19}$
B. Kramer,$^{17,20}$
W. A. Baan,$^{5,21}$
C. Brogan,$^{22}$\newauthor
T. Hunter,$^{22}$
S. Kurtz,$^{23}$
A. M. Sobolev,$^{19}$
J. Brand,$^{24}$
L. Volvach$^{6,11}$
\\
$^{1}$Joint Institute for VLBI ERIC, Oude Hoogeveensedijk 4, 7991 PD Dwingeloo, The Netherlands.\\
$^{2}$Mizusawa VLBI Observatory, National Astronomical Observatory of Japan, 2-21-1 Osawa, Mitaka, Tokyo 181-8588, Japan\\
$^{3}$Korea Astronomy and Space Science Institute, 776 Daedeokdae-ro, Yuseong-gu, Daejeon 34055, Republic of Korea\\
$^{4}$School of Natural Sciences, University of Tasmania, Private Bag
37, Hobart, Tasmania 7001, Australia\\
$^{5}$Xinjiang Astronomical Observatory, Chinese Academy of Sciences,
150 Science 1-Street, Urumqi, Xinjiang 830011, China\\
$^{6}$Astro Space Center, Lebedev Physical Institute, Russian Academy of Sciences, Leninskiy Prospekt 53, Moscow 119333, Russia\\
$^{7}$INAF Osservatorio Astronomico di Cagliari, Via della Scienza 5, 09047 Selargius, Italy\\
$^{8}$Centre for Astronomy, Faculty of Physics, Astronomy and Informatics, Nicolaus Copernicus University, Grudziadzka 5, 87-100 Toru\'n, Poland\\
$^{9}$ The University of Western Ontario, 1151 Richmond Street. London, ON N6A 3K7, Canada\\
$^{10}$Hartebeesthoek Radio Astronomy Observatory, PO Box 443, Krugersdorp 1740, South Africa\\
$^{11}$Radio Astronomy and Geodinamics Department of Crimean Astrophysical Observatory, Katsively, RT-22 Crimea\\
$^{12}$Lomonosov Moscow State University, Sternberg Astronomical Institute, Moscow 119234, Russia\\
$^{13}$National Radio Astronomy Observatory, P.O. Box O, 1003 Lopezville Rd., Socorro, NM 87801\\
$^{14}$Sterrewacht Leiden, Leiden University, Postbus 9513, 2300 RA Leiden, the Netherlands\\
$^{15}$South African Radio Astronomy Observatory (SARAO), 3rd Floor, The Park, Park Road, Pinelands, Cape Town, 7405, South Africa\\
$^{16}$Space Research Unit, Physics Department, North West University, Potchefstroom 2520, South Africa\\
$^{17}$National Astronomical Research Institute of Thailand, 260 M.4, T. Donkaew, Amphur Maerim, Chiang Mai, 50180, Thailand\\
$^{18}$University of Science and Technology, Korea (UST), 217 Gajeong-ro, Yuseong-gu, Daejeon 34113, Republic of Korea\\
$^{19}$ Astronomical Observatory, Ural Federal University, Lenin Ave. 51, Ekaterinburg 620083, Russia\\
$^{20}$Max-Planck-Institut f{\"u}r Radioastronomie, Auf dem H{\"u}gel 69,
53121 Bonn, Germany\\
$^{21}$Netherlands Institute for Radio Astronomy, Oude Hoogeveensedijk 4, 7991 PD Dwingeloo\\
$^{22}$NRAO, 520 Edgemont Road, Charlottesville, VA 22903, USA\\
$^{23}$Instituto de Radioastronom{\'\i}a y Astrof{\'\i}sica, Universidad Nacional Aut$\acute{o}$noma de M$\acute{e}$xico, Apartado Postal 3-72, Morelia 58089, M$\acute{e}$xico\\
$^{24}$INAF-Istituto di Radioastronomia and Italian ALMA Regional Centre, via P. Gobetti 101, 40129, Bologna, Italy\\
}

\date{Accepted XXX. Received YYY; in original form ZZZ}

\pubyear{2018}

\begin{document}
\label{firstpage}
\pagerange{\pageref{firstpage}--\pageref{lastpage}}
\maketitle

\begin{abstract}
This paper reports observations of a 22 GHz water maser `superburst' in the G25.65+1.05 massive star forming region, conducted in response to an alert from the Maser Monitoring Organisation (M2O). Very long baseline interferometry (VLBI) observations using the European VLBI Network (EVN) recorded a maser flux density of $1.2 \times 10^{4}$ Jy. The superburst was investigated in the spectral, structural and temporal domains and its cause was determined to be an increase in maser path length generated by the superposition of multiple maser emitting regions aligning in the line of sight to the observer. This conclusion was based on the location of the bursting maser in the context of the star forming region, its complex structure, and its rapid onset and decay.
\end{abstract}




\section{Introduction}



Galactic maser emission is predominantly associated with the births and deaths of stars. At radio frequencies their uses as astrophysical tools are numerous; revealing milliarcsecond-scale structures and three dimensional motions in the smallest, densest regions of activity \citep{Moscadelli16a, Moscadelli18a} which are typically inaccessible to other wavelengths. Despite a heavy reliance on their functionality, the mechanisms of certain maser behaviours remain to be fully explained, particularly their temporal flux variability.

`Maser burst' and `maser flare' are both terms that have been used in recent literature to describe a sudden increase in the intensity of maser emission. We emphasise that the term `maser burst' in this manuscript is employed to this same meaning and does not refer to an accretion burst, or any other physical `burst'. The term `superburst' refers to a particularly extreme class of maser bursts, where a maser emitting region exhibits a sudden increase in flux density of several orders of magnitude. There are 3 recognised water maser superburst star forming regions (SFRs): Orion KL \citep{Matveenko88,Garay89,Shimoikura05,Hirota11,Hirota14b}, W49N \citep{Honma04,Kramer18a,Volvach19c} and G25.65+1.05 \citep{Volvach17ATEL,Ashimbaeva17,Volvach19}. Sudden enhancement in maser flux density, though less extreme, has also been seen in species other than water \citep{Gordon18,Szymczak18} and in evolved stars \citep{Espiov99,Vlemmings14,Gomez15,Etoka17}.

Superbursts are both rare and transient in nature and consequently their mechanism of action remains enigmatic. 
The widely cited \citet{DnW89} model generally allows for three avenues of flux enhancement of maser emission: an increase in the incident continuum photon flux being amplified; mechanical/radiative induction of more favourable maser pumping conditions; or an increase in the path length of the maser cloud along the line of sight to the observer. 


These different burst scenarios associate specifically to a variety of spatial and physical scenarios which help to place constraints on our understanding of the local medium.
Superbursts therefore represent an avenue to investigate the nature of maser emission under atypical circumstances, exposing amplification conditions.


G25.65+1.05, hereafter G25, is a region of massive star formation which recently achieved notoriety due to its recurring maser burst behaviour, exhibiting kJy bursts on several occasions \citep{Lekht18,Volvach19,Volvach19b}. The source distance is disputed, with estimates ranging from 2.7 kpc \citep{Sunada07} to 12.5 kpc \citep{Green11}, with the latter value making G25 the most powerful maser in the Galaxy \citep{Volvach19b}



In this paper we introduce new VLBI observations targeting the superburst activity of water maser emission in G25. We investigate the bursting mechanism by structural analyses of individual maser features; the overall distribution of masers in the context of the star forming region; and maser temporal behaviour via cross-matching VLBI spatial maser features with spectral maser features from historic and recent single-dish monitoring campaigns. Our discussion builds on the context set out in VLA observations made on the 9th December 2017, described in \citet{Bayandina19}, whose work serves as a basis for forthcoming studies of the G25 water maser.


\begin{figure}
\begin{center}
\includegraphics[width=0.47\textwidth]{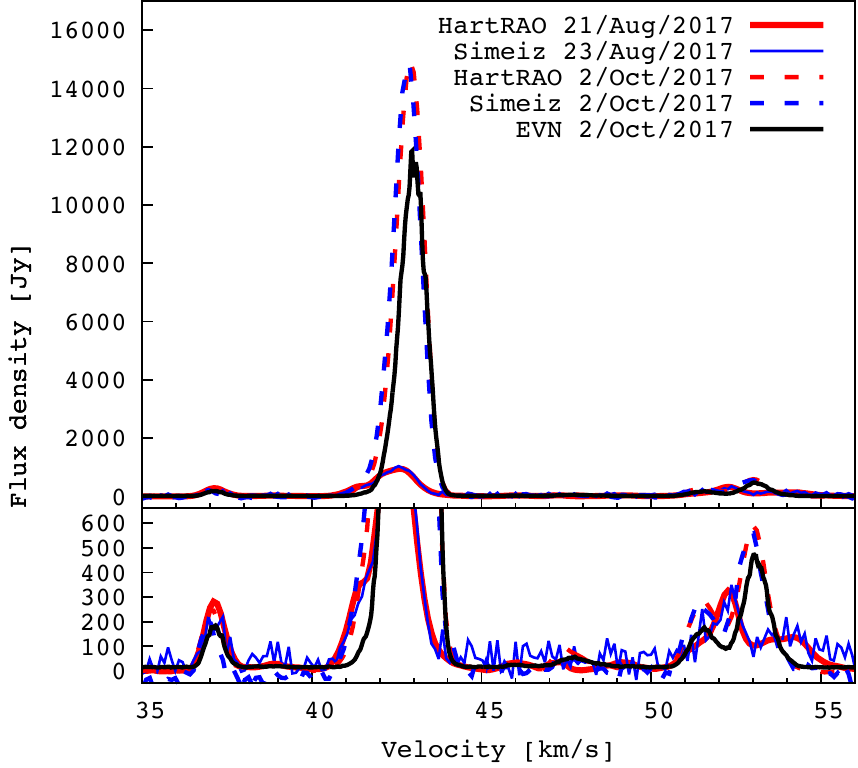}
\caption{The scalar averaged, cross-power water maser spectrum measured with the EVN (solid, black line) is shown in comparison to single-dish spectra provided by the M2O on the same date (dashed red and blue lines), and (solid red and blue lines) pre-flare. The panel below highlights the low-flux density maser features.
\label{SDSPECTRA}}
\end{center}
\end{figure}

\begin{figure*} 
\begin{center}
\includegraphics[width=\textwidth]{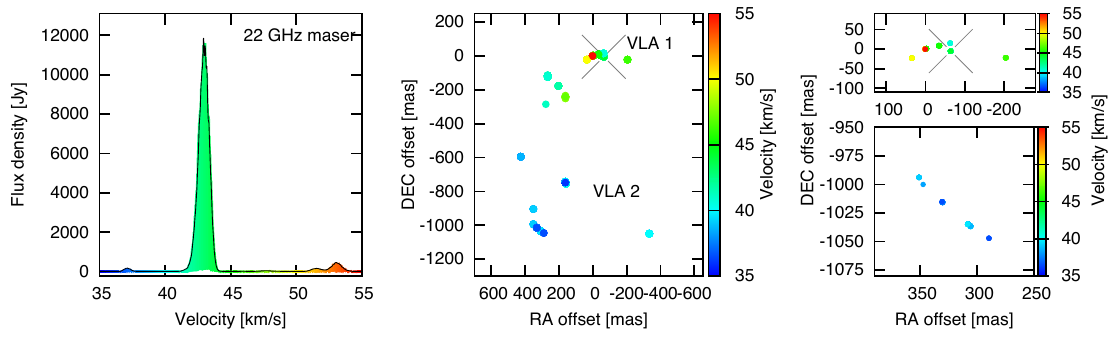}
\caption{\emph{Left} shows the scalar averaged spectrum of maser emission in G25. \emph{Middle} shows the spatial distribution of maser spots in G25 and \emph{right} shows blow-ups of the arc structures near VLA 1 and VLA 2, above and below, respectively. The superburst maser is identified with a cross. In all sub-plots colours indicate LSR velocity.
\label{SPOTMAP}}
\end{center}
\end{figure*}

\section{Observations and data reduction}

Target of Opportunity (ToO) observations of the water maser in G25 were requested to the European VLBI network (EVN) in response to the September 2017 burst activity reported by the single-dish Maser Monitoring Organisation (M2O\footnote{MaserMonitoring.org}, a global co-operative of maser monitoring programs). Six stations were able to respond to the ToO request: Effelsberg, Jodrell Bank (MkII), Onsala (20 m), Toru\'n, Yebes and Hartebeesthoek. Observations were carried out on the 2nd of October 2017 in eVLBI mode whereby locally timestamped, DBBC filtered data are transferred directly to the SFXC correlator at JIVE \citep{Keimpema15}. This approach circumvents data shipping delays and thus enables quicker evaluation of transient events.

Data were transferred at 128 Mbps, comprising one single 16 MHz frequency band with 2-bit, Nyquist sampling and dual circular polarisations. Bands were centred at the rest frequency of the water maser, 22.235080 GHz. Correlated data were generated with 8192 frequency channels to provide 2 kHz channel separation, equivalent to 0.03 km s$^{-1}$. 


Using this setup, the array observed M87 for one hour preceding the main science programme in order to synchronise the eVLBI data streams of participating stations to within $\sim 16~ \mu$s. G25 was observed in 30 minute scans, breaking at the 12 and 18 minute marks for Tsys calibration of $<15$ s duration.
At each hour J2202+4216 (BL Lac) was observed as a delay, bandpass and polarisation calibrator. We also included two 10 min scans each of DA553 and 4C075 for electric vector position angle (EVPA) calibration and check. W49N and J1905+0952 were also observed - these data, and the polarisation results for G25, will be presented in forthcoming works. The scheduled observing time was 14 hrs, of which G25 was scheduled for 3.26 hours.

Fringes to all stations were obtained except for the left circular polarisation data from Toru\'n. An issue with the VLBI firmware affecting the last EVN sessions of 2017 and first session of 2018 lead to erroneous amplitudes across participating stations. We discuss efforts to circumvent this issue below.
Data products were processed to FITS files at JIVE\footnote{Joint institute for VLBI ERIC, www.jive.nl} and are publicly available from the EVN archive under observation code RB004.

\subsection{Data reduction}
Primary calibration was carried out using the AIPS\footnote{Astronomical Image Processing System, www.aips.nrao.edu} software package. Station based flags, bandpass and gain calibration tables produced by the EVN data reduction pipeline were sourced from the online archive and appended to the main data set. These calibration tables also include a-priori ionospheric, correlator sampler, and parallactic angle corrections. Delay and bandpass solutions were derived using J2202+4216 and applied to all sources. The aforementioned amplitude issue was circumvented by scaling the observed flux density of J2202+4216 based on independent VLBI K-band measurements from the GENJI monitoring project \citep{Nagai13}.

Dispersive phase and rate variations, dominated by tropospheric fluctuations, were corrected by fringe fitting on a reference maser with a peak flux density of 409 Jy and simple, unresolved emission, inferred from its constant flux density as a function of projected baseline length.

\begin{figure*} 
\begin{center}
\includegraphics[width=0.75\textwidth]{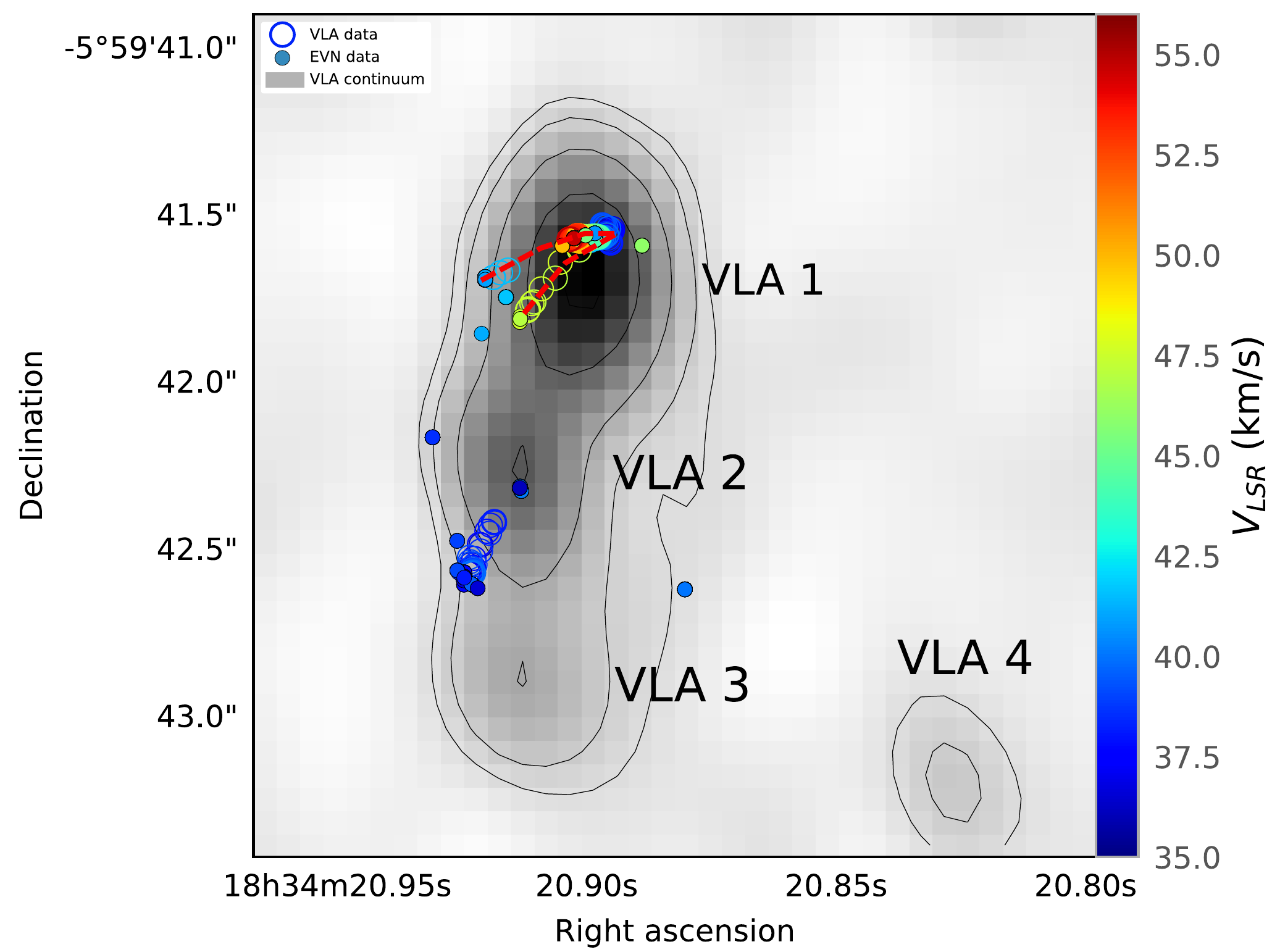}
\caption{Visual summary of the water maser and 22 GHz continuum emission in G25, comparing results from this work with VLA data from \citet{Bayandina19} taken on the 9th December 2017, from which the continuum sources VLA 1, 2, 3 and 4 are labelled. Open circles indicate maser emission detected by the VLA while filled circles represent masers from this work. Colours indicate line of sight velocity and the red dashed line delineate the lateral `V' discussed in Section~\ref{CONC}. \label{EVN_VLA}}
\end{center}
\end{figure*}




After applying all phase and gain calibrations to the full spectral line data set, small shifts were made in the frequency domain to account for time-dependent Doppler shifts caused by Earth rotation. Finally an image cube was produced for spectral line channels in a $+35$ to $+55$ km s$^{-1}$ velocity range, covering a region of $1.4 ^{\prime \prime} \times 1.4 ^{\prime \prime}$ based on the VLA maser maps of \citet{Bayandina19}. The synthesised beam had dimensions of $1.2 \times 0.7$ milliarcseconds, with a position angle of $81 \degree$ measured anti-clockwise from North. The similarity of the cross-power spectrum evaluated within the primary beam (POSSM) with a spectrum evaluated using only the emission in our mapped region (ISPEC) indicated that most of the emission had been successfully included in the map.

The final image cube was searched for emission exceeding a signal-to-noise cutoff of 7 for the majority of the map, however, a higher cutoff value of 100 was necessary for bright maser channels with elevated image noise due to dynamic range limiting.
Henceforth, a maser `spot' refers to emission, per channel, originating from a maser cloud.
All flux densities discussed in this paper, including single dish measurements, are subject to a 10\% absolute calibration uncertainty.


\section{Results}

\subsection{Spectral distribution of masers in G25}
The 22 GHz water maser spectrum measured by the EVN is shown in Figure~\ref{SDSPECTRA} (black solid line). Masers were detected in the range of $+35 < v_{LSR} < +55$ km s$^{-1}$ with the bursting maser occurring at $v_{LSR} = +42.9$ km s$^{-1}$, i.e. close to the central velocity of the spectral features. 
Molecular line observations place the systemic velocity of G25 at $v_{LSR} \sim +42$ km s$^{-1}$ \citep{Mol96,Bronfman96} thus masers in G25 exhibit only modest velocities in the line of sight direction.

Figure~\ref{SDSPECTRA} also shows single-dish spectra during- and prior to the September 2017 superburst; the EVN detected all spectral features in G25 and the number of spectral components remains constant in time. Significant temporal changes in flux were exclusive to the bursting feature which increased from $10^2$ Jy in the quiescent phase, to $10^4$ Jy in the superburst phase. In contrast to the superburst feature, the profile of the feature at $+37$ km s$^{-1}$ remained constant across all observations while only small changes were seen in the $+50$ to $+55$ km s$^{-1}$ range, as is shown in the lower panel of Figure~\ref{SDSPECTRA}.


\subsection{Spatial distribution of masers in the G25 SFR}

The VLBI maser spotmap is shown in Figure~\ref{SPOTMAP}.
The majority of maser emission in G25 resides in the form of two arc structures: an E-W orientated arc near \emph{(x,y)=(0,0) mas offset} which contains the bursting feature at the arc centre, and a NE-SW arc near \emph{(x,y)=(350,-1000) mas offset} (Figure~\ref{SPOTMAP}, \emph{right}). 
These arcs coincide with the two brightest radio continuum sources in G25; VLA 1 and VLA 2 (see Figure~\ref{EVN_VLA}). We henceforth adopt these names for the maser arcs in relevant discussion below. Masers associated with VLA 1 exhibit the full range of velocity components, $+35$ to $+55$ km s$^{-1}$, while masers associated with VLA 2 were typically in the range of $+35$ to $+40$ km s$^{-1}$. Maser spots were also observed in the region between VLA 1 and 2, these masers were closer in velocity to the VLA 2 group.



\subsection{Flux density and the structure of masers in G25}



The highest flux density measured by the EVN was $1.2 \times 10^{4}$ Jy (Figure~\ref{SDSPECTRA}; Figure~\ref{SPOTMAP}, \emph{Left}). Figure~\ref{SDSPECTRA} shows a comparison of the EVN cross-correlation flux density, which is sensitive only to the milliarcsecond scale, to spectra measured independently by single-dish instruments at the Hartebeesthoek Radio Astronomy Observatory (HartRAO) and Simeiz Radio observatory \citep{Volvach19} which are sensitive to emission on all angular scales within the primary beam (a few arcminutes). A comparison of the peak flux densities reported by HartRAO, $1.5 \times 10 ^4$ Jy, and Semiez, $1.5 \times 10 ^4$ Jy, with that of the EVN, $1.2 \times 10 ^4$ Jy, reveals that $80 \pm 10$\% of the bursting maser flux emanates from the milliarcsecond scale, i.e. the superburst maser emission in G25 is highly compact. Similarly high percentage flux recovery was observed for non-bursting maser features indicating that most masers in G25 are highly compact. The EVN did not completely recover all maser emission from the `blue' side of bursting feature (around $+41$ km s$^{-1}$, Figure~\ref{SDSPECTRA}) indicating that the emission is extended.



\begin{figure*}
\begin{tabular}{ccc}
 \hspace{-0.27cm}
 \includegraphics[valign=T,width=0.215\textwidth]{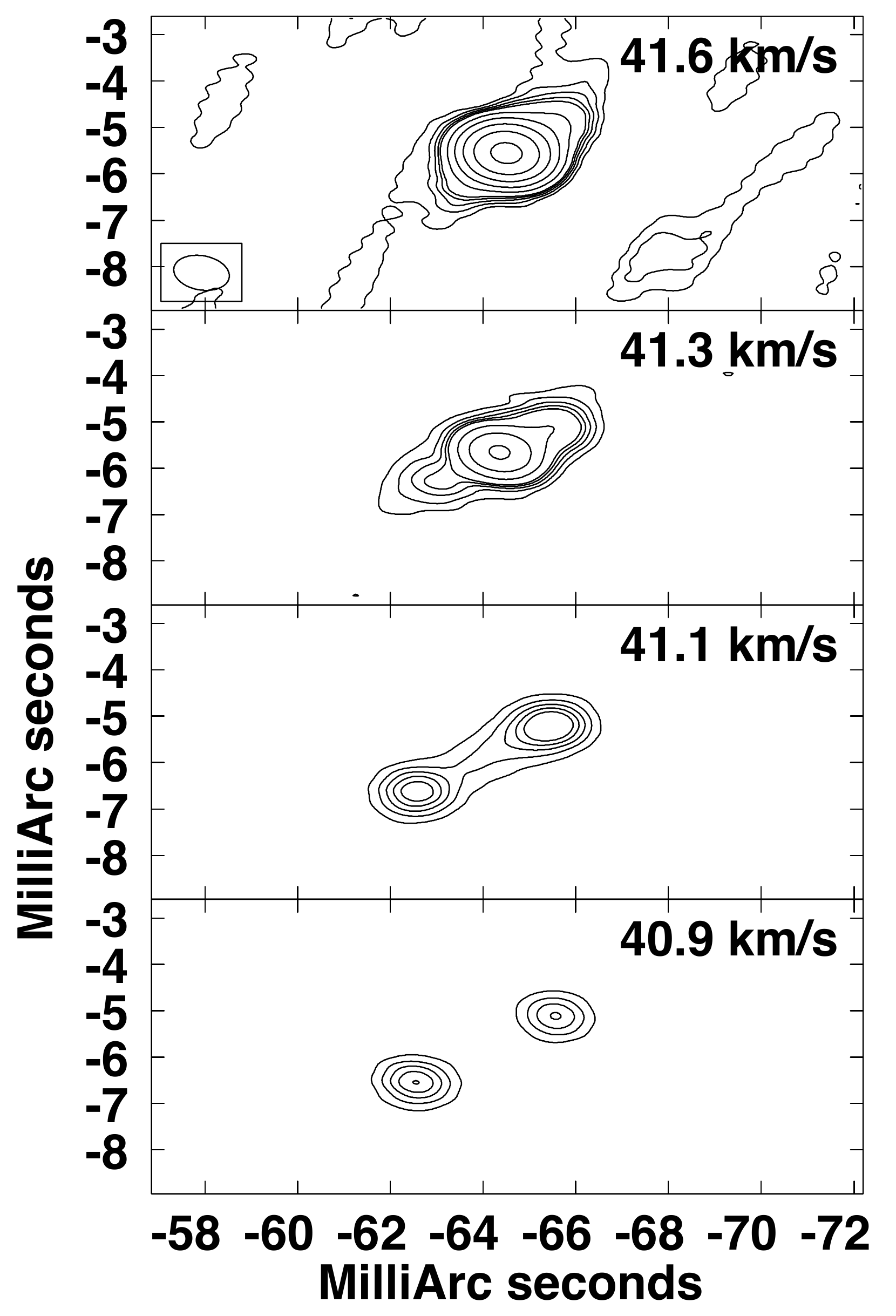} &    \includegraphics[valign=T,width=0.30\textwidth]{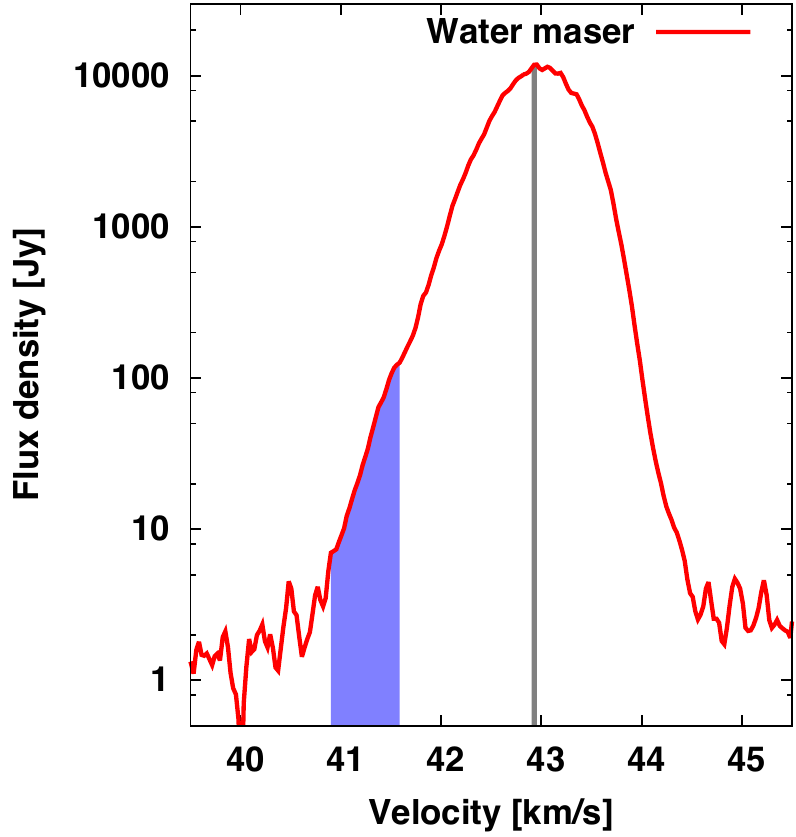} &
 \includegraphics[valign=T,width=0.44\textwidth]{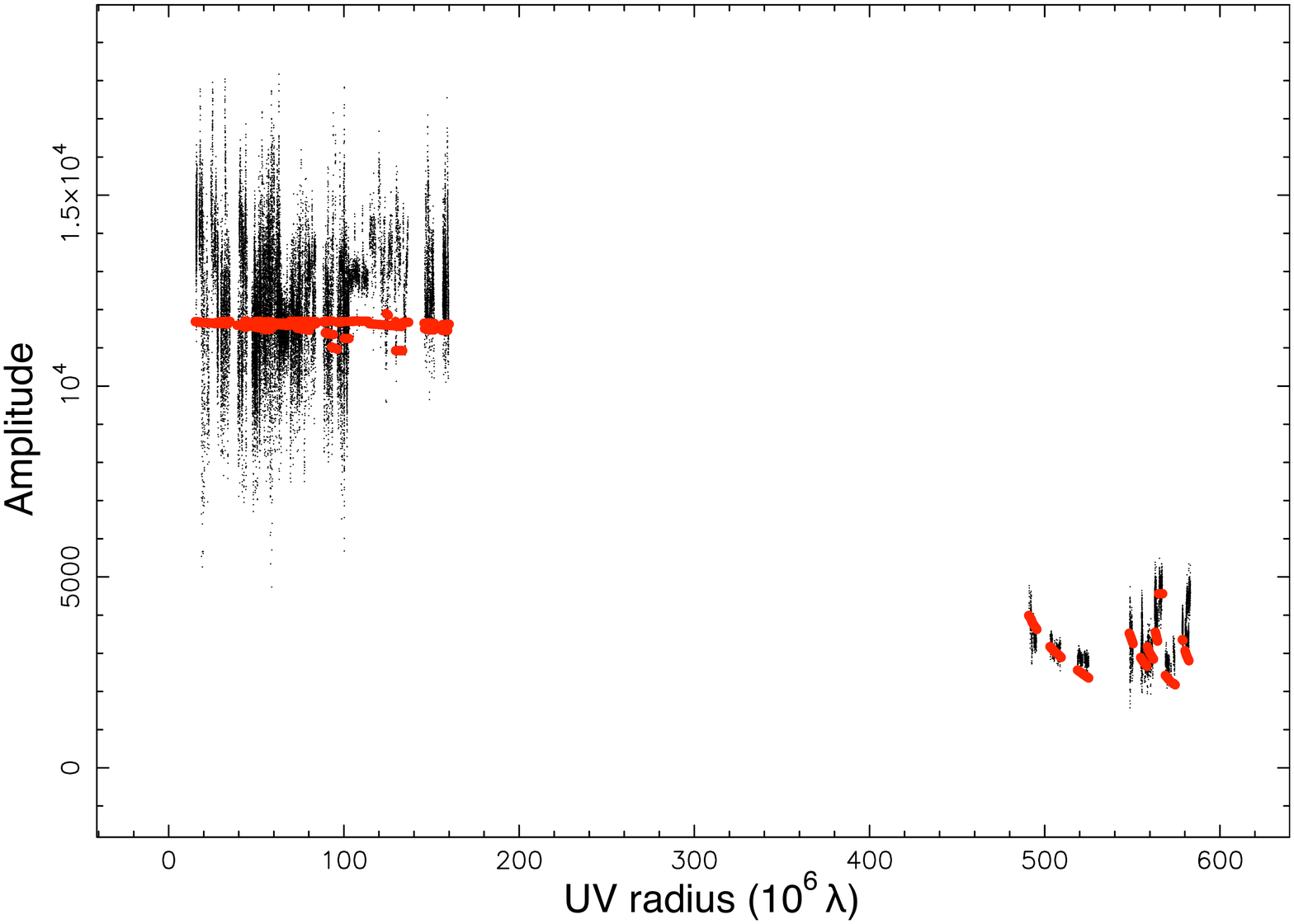}
\end{tabular}
\caption{(\emph{Center}) A logarithmic scale spectrum highlighting the bursting maser feature, in which the peak channel is indicated with a grey line and the blue wing is coloured. (\emph{Left}) Channel maps of the blue wing convey the structure of the emission with contours at 1, 2, 3, 4, 5, 10, 20, 40, 80, 160 multiples of 0.5 Jy which is a typical $3\sigma$ noise value in the off-peak channels. Coordinates are relative to the reference maser spot. (\emph{Right}) A plot of amplitude as a function of projected baseline length for the peak maser channel (black) and a linear model with a Gaussian flux density profile (red) indicating structure at sub-/milliarcsecond scale. \label{MOMNT}}
\end{figure*}

Two kinds of complex structure were found in the EVN observations of the bursting spectral component. Firstly, emission in the brightest spectral channel exhibited two-scale structure comprising both compact and an extended emission (Figure~\ref{MOMNT} \emph{right}). Secondly, when imaging the lower flux channels of the superburst spectral feature, the maser emission was seen to decompose into two spatially distinct peaks in the blue wing limit (Figure~\ref{MOMNT} \emph{left}). Each of these characteristics are discussed in the following subsections.

\subsection{Compact and extended emission}
Structural analysis of maser emission in interferometric data is possible via consideration of the flux density as a function of projected baseline length. A constant visibility amplitude with increasing baseline length in such a plot indicates emission that is unresolved on all baselines, as was the case for the reference maser on which the phase calibration was based. On the other hand, a decreasing visibility amplitude with baseline length indicates that emission is partially resolved-out on the longer baselines. 
The lack of intermediate length baselines between the continental and intercontinental EVN stations hinders an in-depth structural analysis, however, some basic assertions can be made.
Figure~\ref{MOMNT} (\emph{right}) reveals that the superburst maser component in G25 is partially resolved, with emission on two spatial scales; a milliarcsecond-scale component of $\sim1.0\times 10^4$ Jy and a sub-milliarcsecond component of $\sim4.0\times 10^3$ Jy. These correspond to brightness temperature lower limits of $4.7\times10^{13}$ K and $2.0\times10^{14}$ K, respectively.

The visibility data sampled at the peak flux channel were best fit by a linear structure whose flux profile is a Gaussian of 0.3 milliarcsecond full width at half maximum, at a position angle of PA $= 115^{\circ}$. Note that the elongation of emission is markedly different from the PA of the synthesised beam ($81 \degree$), ruling out beam effects.
The maser data and linear model are shown as functions of projected baseline length in Figure~\ref{MOMNT} (\emph{right}).

\subsection{Decomposition of the blue wing}

The spectral profile of the bursting maser feature revealed significant excess flux in the blue wing near $+41$ km s$^{-1}$, which is especially visible when viewed in log scale in Figure~\ref{MOMNT} (\emph{middle}). The channels comprising this blue wing were imaged and the corresponding channel maps are shown in Figure~\ref{MOMNT} (\emph{left}). The maps reveal complex structure in the blue wing of the G25 superburst maser feature delineating a clear decomposition into two spatially distinct maser features at a NW-SE position angle, specifically PA $\simeq 115^{\circ}$.



\section{Discussion}

For in-depth discussions on astrophysical masers the reader may refer to \citet{DnW89,Elitzur91,Gray99,Gray12,Sobolev12}. We only list here some of the key properties of maser emission that should be considered when interpreting the maser superburst presented in this paper.

\subsection{Possible routes of the superburst mechanism}

Masers are the product of stimulated emission in which an incident photon (from the local environment or from background radiation) initiates an exponentially proliferating de-excitation cascade; each input photon leading to two output photons for each occurrence of stimulated emission. This behaviour is sustained providing the existence of a sufficient pool of excited molecules to de-excite, i.e. the rate of populating the excited
state(s) must be faster than the rate of de-excitation; the maser is then said to be `unsaturated'. When de-excitation surpasses excitation rates the maser can no longer amplify with an exponential relation to the maser path length and the maser is said to be `saturated'.

The three key factors influencing the observed maser output are therefore the intensity of incident emission, the abundance of population-inverted molecules, and the number of de-excitable molecules along the observer's line of sight.
As such, these form the three general routes considered in explaining the superburst mechanism. Discerning which of these factors is responsible for the superburst observed in G25 is possible by noting some distinguishable observational signatures. 




In the case of an increase in incident photons, all maser emitting gas (of any molecular species) foreground to the source of continuum emission would exhibit an enhanced flux. Masers, either compact or extended, covering a wide range of velocities may be involved, as was seen in NGC6334I \citep{Gordon18}. Temporally, the maser enhancement should reflect changes in the radiation field where, in the unsaturated domain, a linear increase would lead to an exponential maser flux density increase (as was seen in \citealt{Szymczak18}). In the cases of S255IR and NGC6334I several radiation driven maser transitions were enhanced for several weeks while some transitions remained enhanced for several years \citep{Szymczak18,Gordon18} suggestive of a lasting elevated local continuum radiation field.


In the case of water masers, enhanced pumping conditions in a shock could contribute the collisional energy required for producing population inversion of the 22 GHz water maser transition \citep{Hollenbach13}. Observationally, this would manifest as enhanced flux only for maser emitting regions proximal in location to the shocked gas, which, for proto-/stellar jets, are typically several 100 AU (e.g. \citealt{Goddi11,Cesaroni18}). Temporally, water maser bursts associated with new shocks would depend on the balance of pumping conditions.


Occurrences of stimulated emission increase along the propagation path of the maser gas. This can occur in the case of a geometrically or structurally variable maser complex. \citet{Gray19} show that in the case of a rotating maser cloud, quasi-periodic maser variability can be produced, however strong flaring behaviour cannot. Conversely, a significant increase in path length can also be achieved when multiple maser emitting regions producing emission at the same frequency overlap on the sky-plane viewed by the observer; an effect confined to a very narrow spectral domain but not requiring association with shocks or continuum regions (e.g. \citealt{Shimoikura05}). Temporally, the burst would persist as long as the alignment is maintained which would depend on the angular scale and relative motions of the individual components as viewed from the perspective of the observer. While one superburst in Orion KL lasting several hundred days was found to be associated with two overlapping maser emitting regions \citep{Shimoikura05}, far shorter bursts of less than 2 days duration are known to have occurred in the same source \citep{Matveenko88}. Assuming typical Galactic proper motions of several milliarcseconds/year a maser overlap region on the order of 1/365 milliarcseconds would be required to achieve temporal variability on an intra-day level.


\subsection{The superburst in G25}
\label{BURST}

Attributing the superburst in G25 to one of the aforementioned scenarios is facilitated by considering the available observational evidence; namely, the maser structure and distribution in the context of the G25 SFR, and temporal behaviours known from historic monitoring campaigns \citep{Lekht18,Volvach19}. These are each discussed below. It should also be noted that despite the high flux densities involved the masers in G25 remain unsaturated \citep{Volvach19,Volvach19b}.

\subsubsection{Spatial and environmental considerations}

G25 contains four centimeter sources, VLA 1, 2, 3 and 4, the natures of which are discussed in \citet{Bayandina19}. VLA 2 is the only detected millimeter source in the G25 SFR and also houses class II methanol masers. It is thus likely to represent the embedded massive star which drives activity in this region. The superburst maser is located in VLA 1 (See Figure~\ref{EVN_VLA}, edited from \citealt{Bayandina19}) which is the brightest source of 5 cm emission in the G25 region, and, in contrast to VLA 2, exhibits no class II methanol masers or millimeter emission. VLA 1 may therefore constitute a shock region which would be consistent with the production of collisionally excited water masers.


Shocks produced in bipolar ejections from massive stars exhibit typical sizes of $\sim 300$ AU (eg. \citealt{Goddi11,Cesaroni18}) and protostellar disks exhibit similar scales \citep{Simon00,Hirota14,Carpenter16}. In contrast, almost all of the flux density enhancement in the G25 superburst originated from a single sub-milliarcsecond region, which is $<2.7$ to $<12.5$ AU, depending on the adopted distance (2.7 kpc or 12.5 kpc).  The largest maser structure is the arc associated with VLA 1 which hosts the superburst maser feature (Figure~\ref{EVN_VLA}). Its 300 milliarcsecond extent would correspond to 810 or 3750 AU depending on adoption of the near or far distance, respectively. If the maser burst were driven by new shock energy the confinement of the flux enhancement to one part of the arc becomes difficult to explain.




\subsubsection{Temporal considerations}

We can infer the historic temporal behaviour of the bursting maser feature by following the $+42.8$ km s$^{-1}$ maser emission of \citet{Volvach19} (see their Figure 2).
Temporal variations in the maser flux in G25 are some of the most extreme reported in the literature, with three extreme flares occurring between 2002 and 2016 \citep{Lekht18}, including several short, and even intra-day bursts \citep{Ashimbaeva17,Volvach19}. Such rapid increase and subsequent decrease timescales are much faster than those reported in association with shocks and accretion events \citep{Gordon18,Szymczak18}. As mentioned previously, assuming $\sim$ mas yr$^{-1}$ proper motions the intra-day variability seen in G25 implies an overlap region of angular size of the order 1/365 milliarcseconds. Such a compact source structure in the superburst component is supported observationally in G25 (see below).

\subsubsection{Structure of the superburst water maser in G25}

While a comparison of single-dish and VLBI scale maser spectra (Figure~\ref{SDSPECTRA}) reveal that $80 \pm 10 $\% of the total maser emission in G25 emanates from the miliarcsecond scale, a comparison of continental and intercontinental EVN data for the $+42.9$ km s$^{-1}$ channel reveals contributions to flux density on the milliarcsecond and sub-milliarcsecond scales (Figure~\ref{MOMNT} \emph{right}). The detection of the same maser feature on space VLBI baselines \citep{Kurtz18,Bayandina19b} confirms that the bursting maser feature in G25 is extremely compact.

VLBI imaging of the emission uncovered an elongation in the NW-SE direction, which was best fit by a model of a linear extension with a Gaussian flux density profile (Figure~\ref{MOMNT} \emph{right}). While a map of the emission in the blue wing of the flaring feature (around $+41$ km s$^{-1}$) revealed two spatially distinct peaks which flank the location of the flaring feature (Figure~\ref{MOMNT} \emph{left}). The NW-SE orientation of the spatially separated peaks matches the position angle of the aforementioned linear model fit to the peak channel which suggests causality, implying that the superburst feature and two weaker features are associated.

\section{Conclusion and hypothesis}
\label{CONC}

The structure and flux density profile in the bursting feature in G25 can be explained by a scenario whereby two milliarcsecond-scale maser emitting regions partially overlap on the sky plane at a NW-SE orientation, (See Figure~\ref{MOMNT} \emph{left}). The partial overlap forms a sub-milliarcsecond region of increased maser path length with an extension in the NW-SE as is described by the model fit to the peak maser channel (Figure~\ref{MOMNT} \emph{right}). The small angular size of this region is corroborated by detection of the same feature in space-VLBI observations during the G25 superburst event, as reported by \citet{Bayandina19b}.

Here we hypothesise a scenario whereby such an overlap might occur, and why they are not more commonly observed in maser bearing star forming regions. When viewing the EVN data in the context of VLA data from \citet{Bayandina19}, maser emission associated with VLA 1 has the morphology of a lateral `V' shape, as can be seen in Figure~\ref{EVN_VLA}. The maser superburst occurred in a feature residing at the apex of the `V'. We hypothesise that the lateral `V' traces two linear maser associations, or `sheets' which, from the viewpoint of the observer, intersect at a position in the skyplane corresponding to the position on the superburst maser. This overlap of maser sheets could provide an increase in the path length of maser gas as chance superpositions occur between masers in the two sheets. To achieve a colocation of \emph{velocity coherent} maser sheets would require an alignment in both position, and line of sight velocity, which may explain why such events are rarely seen.

According to this hypothesis a relative proper motion between the two sheets in a general N-S direction would cause the overlap region to propagate laterally in the sky-plane, like the crosspoint of an opening or closing pair of scissors. This propagation would traverse with a proper motion $\mu_{\rm crosspoint} = \mu_{\rm relative} / \tan{\theta}$ where $\theta$ is the angle between the sheets and $\mu_{\rm relative}$ is their relative proper motion. Water masers in shocks typically move at about $v \sim 20$ km s$^{-1}$ \citep{Burns16b,Burns17a} which at distances of 2.7 and 12.5 kpc, and holding one sheet stationary, corresponds to relative proper a motion of $\mu_{\rm relative} \sim 0.3 - 1.5$ mas yr$^{-1}$. For $\theta \sim 20 \degree$ in VLA 1 we estimate that the crosspoint in the maser sheets would propagate in the skyplane at $\mu_{\rm crosspoint} \sim 0.8 - 4.1$ mas yr$^{-1}$. Masers in G25 typically group on $\sim30$ milliarcsecond scales and also form milliarcsecond scale complexes (Figure~\ref{SPOTMAP} \emph{right}), thus our hypothesis would predict chance alignments to occur multiple times during a months-to-years scale flurry, interspersed with periods of decade-scale quiescence. This hypothesis is remarkably consistent with the long term, high cadence light-curves of \citet{Lekht18}.
Maser cloud trajectories attained via VLBI proper motion observations would provide a test of this hypothesis and possibly allow the prediction of future superbursts.



\section{Summary}

The key points of this paper can be summarised as follows: 


EVN (VLBI) data taken of a $1.2 \times 10^{4}$ Jy water maser superburst in G25.65+1.05 were presented, and supplemented by single-dish data from the Maser Monitoring Organisation (M2O).

The superburst was investigated in the spectral, spatial and temporal domains and the results were used to differentiate between the  mechanisms of action considered feasible by basic maser theory.

The superburst feature presented an elongated structure with PA $= 115^{\circ}$, and was flanked by two weaker, spatially separated features, their relative position angle was also PA $\simeq 115^{\circ}$.

VLBI and single dish data consistently favour a scenario whereby an alignment, or `overlap', of multiple unsaturated maser emitting regions in the sky-plane produced a superburst via a sudden increase in the maser path length along the line of sight to the observer.

Observations locate the bursting maser feature at the intersect of two maser sheets. 
Multi-epoch VLBI measurements of the trajectories of maser features associated with the maser `sheets' could allow the prediction of future maser superbursts in G25.


\section*{Acknowledgements}
RB acknowledges support through the EACOA Fellowship from the East Asian Core Observatories Association.
GO acknowledges support from the Australian Research Council Discovery project DP180101061 funded by the Australian Government, the CAS `Light of West China' Program 2018-XBQNXZ-B-021 and the National Key R\&D Program of China 2018YFA0404602. TH is financially supported by the MEXT/JSPS KAKENHI Grant Number 17K05398. BM acknowledges support from the Spanish Ministerio de Econom\'ia y Competitividad (MINECO) under grants AYA2016-76012-C3-1-P and MDM-2014-0369 of ICCUB (Unidad de Excelencia ``Mar\'ia de Maeztu'’). NSh acknowledges support from Russian Science Foundation grant 18-12-00193. AMS is supported by the Ministry of Science and High Education (the basic part of the State assignment, RK No. AAAA-A17-117030310283-7) and by the Act 211 Government of the Russian Federation, contract No.02.A03.21.0006. JOC acknowledges support by the Italian Ministry of Foreign Affairs and International Cooperation (MAECI Grant Number ZA18GR02) and the South African Department of Science and Technology’s National Research Foundation (DST-NRF Grant Number 113121) as part of the ISARP RADIOSKY2020 Joint Research Scheme. This work was supported by the National Science Centre, Poland through grant 2016/21/B/ST9/01455.
This work was funded by the RFBR, project number 19-29-11005. WAB acknowledges the support from the National Natural Science Foundation of China under grant No.11433008 
and the Chinese Academy of Sciences President's International Fellowship Initiative under grant No. 2019VMA0040.

The European VLBI Network is a joint facility of independent European, African, Asian, and North American radio astronomy institutes. Scientific results from data presented in this publication are derived from the following EVN project code: RB004

We would like to thank the anonymous referee whose direction lead to crucial improvements in the scientific impact of this work, in addition to clarification of the manuscript.

\bibliographystyle{mnras}
\bibliography{Kagoshima} 

\small

\end{document}